# SYSTEM IDENTIFICATION FOR INDOOR CLIMATE CONTROL

A.W.M. (Jos) van Schijndel, P.W.M.H. (Paul) Steskens,
Eindhoven University of Technology
Eindhoven

**ABSTRACT**

The study focuses on the applicability of system identification to identify building and system dynamics for climate control design. The main problem regarding the simulation of the dynamic response of a building using building simulation software is that (1) the simulation of a large complex building is time consuming, and (2) simulation results often lack information regarding fast dynamic behaviour (in the order of seconds), since most software uses a discrete time step, usually fixed to one hour. The first objective is to study the applicability of system identification to reduce computing time for the simulation of large complex buildings. The second objective is to research the applicability of system identification to identify building dynamics based on discrete time data (one hour) for climate control design. The study illustrates that system identification is applicable for the identification of building dynamics with a frequency that is smaller as the maximum sample frequency as used for identification. The research shows that system identification offers good perspectives for the modelling of heat, air and moisture processes in a building. The main advantages of system identification models compared to the modelling of building dynamics using building simulation software are, that (1) the computing time is reduced significantly, and (2) system identification models run in a MATLAB environment, in which many building simulation tools have been developed.

## 1. INTRODUCTION

It is widely accepted that the modelling and simulation of the dynamic response of a building contributes to improve user comfort, to reduce energy consumption and to improve Heating, Ventilation and Air Conditioning (HVAC) performance in buildings. For the past 50 years, a wide variety of building energy simulation programs have been developed, enhanced and are in use throughout the building energy community (Crawly *et al*.(2005)).

Currently, two broad but distinct approaches to modelling the dynamic response of a building are common: a forward (classical) approach and a data-driven (inverse) approach. The objective of the forward approach is to predict output variables of a specified model with known structure and known parameters when subject to specified input variables. To ensure accuracy, models have tended to become increasingly complex. This approach presumes detailed knowledge not only of the various natural phenomena affecting system behaviour, but also of the magnitude of various interactions. The main advantage of this approach is that the system need not be physically built to predict its behaviour. This approach is ideal to preliminary design and analysis stage (ASHRAE (2005)).

An alternative approach is a data-driven modelling approach. The objective of the data-driven approach is to determine a mathematical description of the system and to estimate system parameters in a situation, when input and output variables are known and measured. In contrast to the forward approach, the data-driven approach is relevant when the system has already been built and actual performance data are available for model development and identification. Data-driven modelling often allows identification of system models that are not only simpler to use but also are more accurate predictors of future system performance than forward models. In literature, this data-driven modelling approach is also known as system identification (SI) (ASHRAE (2005)).

Several studies show that system identification is a useful tool for building energy simulation (BES) and the analysis of heat, air and moisture (HAM) processes in buildings. Lowry and Lee (2004), for example, describes the application of the data-driven modelling approach to the thermal response of a conventional office space using data collected from an existing building management system. A similar system identification technique for estimation of the heat dynamics of buildings, based on discrete time building performance data is described by Madsen and Holst (1995). Moreover, Cunningham (2001) used system identification techniques to infer room or building ventilation and moisture release rates from psychometric data. Mechaqrane and Zouak (2004) describe the use of system identification to predict the indoor air temperature of a residential building. As an alternative to the use of test facilities, some researchers have compared their models with theoretical predictions from standard computer simulation software such as TRNSYS (Pape *et. al* (1991)). Pape *et. al* (1991) identify building dynamics based on discrete time results from building simulation software with a time step of one hour. Fast dynamic information regarding system dynamics within an hour is not available.

The main problems regarding the simulation of the dynamic response of a building using building simulation software are that:
- o The simulation of a large complex building using standard building simulation software is time consuming.
- o The simulation results often lack information regarding fast dynamic behaviour (in the order of seconds) of the building, since most software uses a discrete time step. This time step is usually fixed to one hour. However, information about fast dynamic behaviour and transient responses, which often lie within this time step, may be essential for application of an appropriate control strategy such as on/off control.

The paper has two objectives. The first objective is to study the applicability of system identification to reduce computing time for the simulation of large complex buildings. The second objective is to research the applicability of system identification to identify building dynamics based on discrete time data (one hour) for climate control design. The study provides an answer to the central questions:
- o *Is system identification useful to reduce computing time of building energy simulations?*
- o *Can system identification be applied for the identification of building dynamics and climate control design?*

Furthermore, an evaluation of the advantages, disadvantages and limitations of system identification considering climate control is discussed, focusing on accuracy and computing time.

The research methodology consists of five stages. First, a model identification toolbox is selected for model identification of building dynamics. Second, the quality of the results produced by this toolbox is evaluated based on the results described in the paper "*Modelling the passive thermal response of a building using sparse BMS data*" (Lowry and Lee (2004)). Third, the applicability of system identification for internal temperature control is researched based on five case studies. Table 1 presents the input and output data that have been used for the identification and application of the SI models:

I. Identification of a building model based on discrete time data, containing free-floating indoor air temperature and application in a situation with free-floating indoor air temperature (continuous time).
II. Identification of a building model based on discrete time data, containing free-floating indoor air temperature and application in a situation with indoor air temperature control (continuous and discrete time).
III. Identification of a building model based on discrete time data, containing on/off-controlled indoor air temperature and application in a situation with similar set points for heating and cooling (discrete time).
IV. Identification of a building model based on discrete time data, containing on/off-controlled indoor air temperature and application in a situation with similar set points for heating and cooling (continuous time).
V. Identification of a building model based on discrete time data, containing on/off-controlled indoor air temperature and application in a situation with different set points for heating and cooling (discrete time).

Moreover, case study III, IV and V involve a parameter study considering the sensitivity of the set points for heating and cooling with respect to the quality of the developed SI models. Fourth, model identification based on similar simulations of Heat, Air and Moisture (HAM) processes in HAMBase (De Wit (2004)) has been evaluated. Lastly, external data, retrieved from a building performance simulation in ESP-r (Data Model Summary ESP-r Version 9 Series (2001)), has been used to identify building dynamics.

Table 1: Case Studies

| Case Study | Identification | Application |
|---|---|---|
| I | Discrete free-floating indoor air temperature 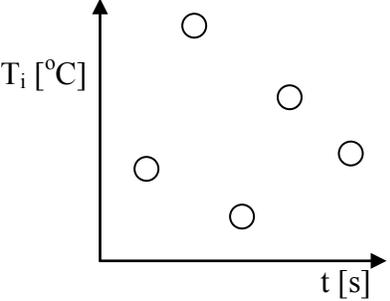 | Continuous free-floating indoor air temperature 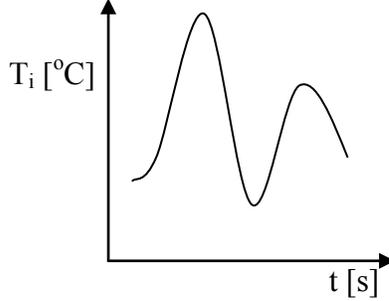 |
| II | Discrete free-floating indoor air temperature 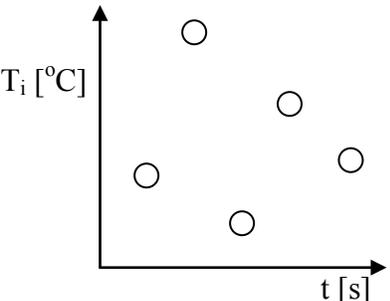 | Continuous and discrete controlled indoor air temperature 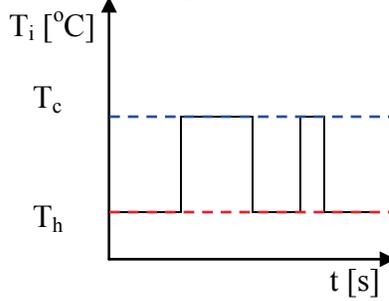 |
| III | On/off controlled discrete indoor air temperature 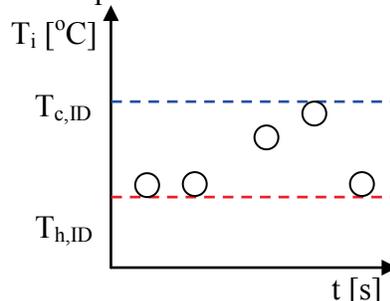 | On/off controlled discrete indoor air temperature 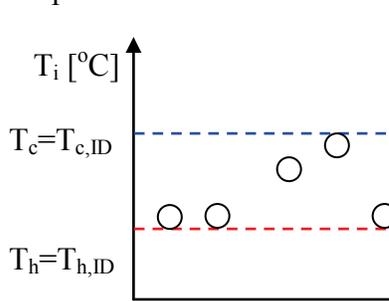 |
| IV | On/off controlled discrete indoor air temperature 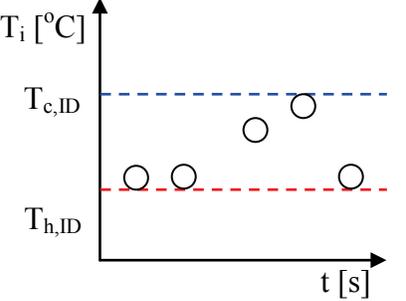 | On/off controlled continuous indoor air temperature 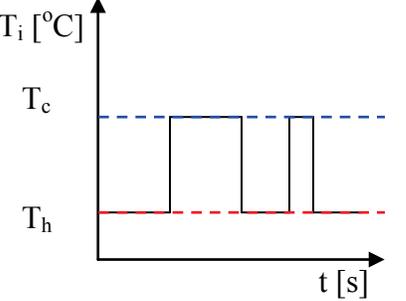 |
| V | On/off controlled discrete indoor air temperature 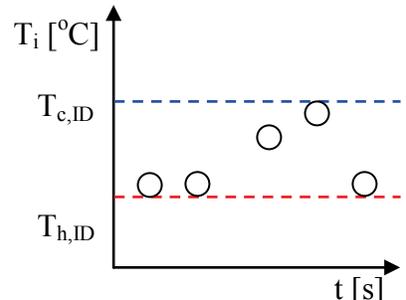 | On/off controlled discrete indoor air temperature 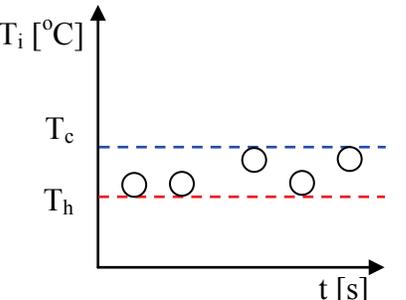 |

The paper is organized as follows: Section 2 describes the verification of the system identification tool, which has been used for model identification. In section 3, the application of system identification based on a HAMBase (De Wit (2004)) building energy simulation is reported. Section 4 reports the application and perspectives of system identification regarding the modelling of HAM transport in buildings. In Section 5 the application of system identification based on an ESP-r building performance simulation is described. Section 6 gives the conclusions and limitations regarding system identification related to the identification of building dynamics.

## 2. LITERATURE BASED VERIFICATION

Lowry and Lee (2004) describe the application of system identification to the thermal response of a conventional office using data collected from an existing building management system (BMS). The input and output data are used for estimation of the system identification model. Data were collected using BMS at 15-min intervals over a three week period, therefore giving more than 2000 measurements. Lowry and Lee (2004) estimate the parameters of an *Output-Error* model (OE211) represented by the Equation (1):

$$T_i = \frac{a.z^{-1} - b.z^{-2}}{c.z^{-1}} T_0 - \frac{d.z^{-1} - ez^{-2}}{f.z^{-1}} p + \varepsilon \qquad (1)$$

The results from the study described by Lowry and Lee (2004) have been reproduced in this research focusing on the verification of the system identification tools. The MATLAB System Identification Toolbox (Ljung (1997)) has been used for model identification. Table 2 shows a comparison of the model coefficients documented in Lowry and Lee (2004) and the coefficients resulting from the present study. Table 1 shows that both models agree well with each other. Moreover, the models are nearly identical.

It is concluded that the MATLAB System Identification Toolbox is an appropriate tool for model estimation. The research proceeds with the identification of building dynamics based on a building energy simulation.

Table 2: Comparison of model coefficients

| Coefficients | a | b | c | d | e | f |
|---|---|---|---|---|---|---|
| Lowry and Lee (2004) | 0.2361 | -0.2334 | 1-0.9961 | -0.03350 | 0.03370 | 1-0.9377 |
| Present study | 0.2346 | -0.2319 | 1-0.9960 | -0.03353 | 0.03364 | 1-0.9374 |

# 3. BUILDING ENERGY SIMULATION MODEL

In this section, the applicability of system identification for indoor air temperature control is researched based on five case studies (Table 1). System identification models have been developed following the methodology that is presented in Figure 1. Three steps have been illustrated: *Identification*, *Simulation* and *Comparison*.

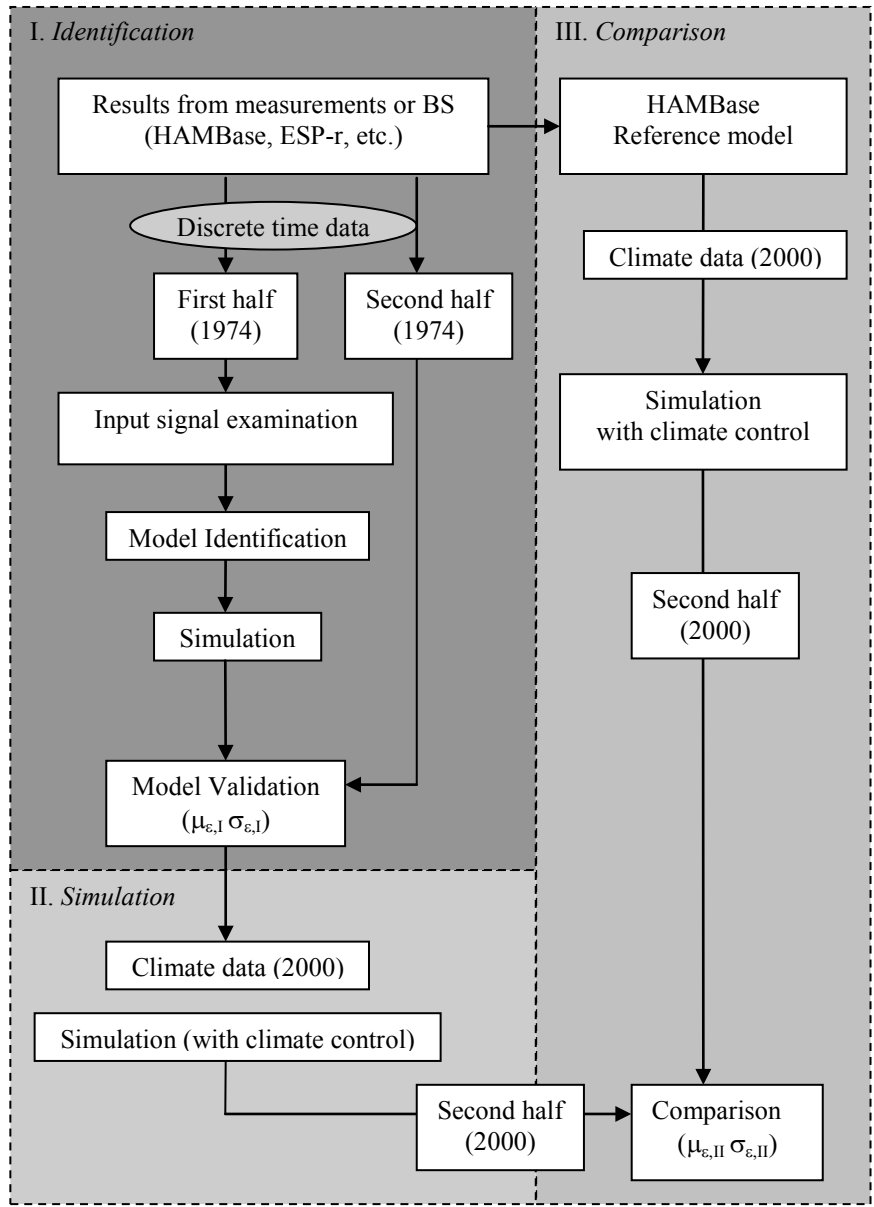

Figure 1: Modelling methodology

The first step is the *Identification* of building dynamics using discrete time data. Discrete time data series are collected from measurements or from building simulation results. The first half of these data series is used for parameter estimation and the second half reserved for model validation. Before the data is used for identification of building dynamics, the input signal is examined.

First of all, the sampling and Nyquist frequency Brigham (2004) of the signal are analyzed. The reader should notice that, when sampling a signal, the sampling frequency must be greater than twice the bandwidth of the input signal in order to be able to reconstruct the original perfectly from the sampled version (Nyquist-Shannon theorem Brigham (2004)). Mathematically, the theorem is formulated as a statement about the Fourier transformation. If a function $s(x)$ has a Fourier transform $F[s(x)] = S(f) = 0$ for $|f| \geq W$ then it is completely determined by giving the value of the function at a series of points spaced $1/(2W)$ apart. The values $s_n = s(n/(2W))$ are called the samples of $s(x)$. The minimum sample frequency that allows reconstruction of the original signal, that is $2W$ samples per unit distance, is known as the Nyquist frequency. With respect to the Nyquist-Shannon theorem, a system identification model is able to capture information regarding building dynamics perfectly for all frequencies with a smaller frequency than the frequency of the data used for identification (Sample frequency). For higher frequencies the identified model may cause deviations.

Second, the input signal should deliver as much input power into the system as possible. The amount of input power, present in the input signal, is defined as the Crest factor $C_f$, Equation (2) (Girod *et al*. (2001)). The smaller the Crest factor, the better the signal excitation resulting in larger total energy delivery and enhanced signal-to-noise ratio. The theoretical lower bound for the Crest factor is 1.

$$C_f = \frac{\max(u(t))}{\sqrt{\langle u^2 \rangle}} \qquad (2)$$

After examination of the input signal, the MATLAB System Identification Toolbox has been used to estimate a state-space model describing building dynamics. The model is validated by comparing the simulation results and the second half of the original discrete time data. The second step is a *Simulation*. The SI model is simulated in MATLAB based on climate data of the year 2000. The third step is the *Comparison* of the results. A reference model of the building is simulated in HAMBase. Both the results from the SI model simulation and the results from the HAMBase reference model are compared.

## 3.1. Free-floating based model identification and free-floating application

This section describes the identification of building dynamics based on a discrete time simulation results with free-floating indoor air temperature. The SI model is applied and simulated in a situation with free-floating indoor air temperature (discrete time). First, an SI model has been identified in accordance with the *Identification* step (Figure 1). Second, the *Simulation* step has been applied (Figure 1). The SI model has been simulated in MATLAB with free-floating indoor air temperature. Third, the results have been compared with the results from the reference HAMBase/Matlab reference model in accordance with the *Comparison* step (Figure 1).

(*I.*) The residential reference building consisting of four rooms has been simulated in HAMBase/Matlab using climate data of the years 1974 and 2000. A simulation time step of one hour is used, giving 8750 values. The temperature in the building is free-floating. The first half of the climate data of the year 1974 has been used for model identification. The second half has been used for model validation. Both model input variables, outside air temperature and solar gains, and output variables, inside air temperature, are presented in Figure 2. Figure 2 illustrates the input and output variables of one room. (The reader should notice that the solar gains are defined as the heat gain supplied to the room by the solar radiance). Before identification of building dynamics, the input signal is examined. The sample frequency of the data is $2.8 \cdot 10^{-4}$ Hz. Based on the Nyquist-Shannon theorem, the model is expected to represent building dynamics for frequencies smaller than $2.8 \cdot 10^{-4}$ Hz. The amount of input power into the system is determined by calculation of the Crest factors $C_{f,T0}$, $C_{f,Qsolar}$, and $C_{f,Ti}$ (Equation (2)). Regarding the outside air temperature and the solar gains the Crest factors are respectively 3.09 ($C_{f,T0}$) and 5.47 ($C_{f,Qsolar}$). Considering the inside air temperature a Crest factor of 2.61 ($C_{f,Ti}$) has been calculated. After examination of the input signal, the MATLAB System Identification Toolbox has been used to develop a state-space model describing free-floating indoor air temperature. Figure 3 shows a representation of the state-space model, which describes the indoor air temperature in the building.

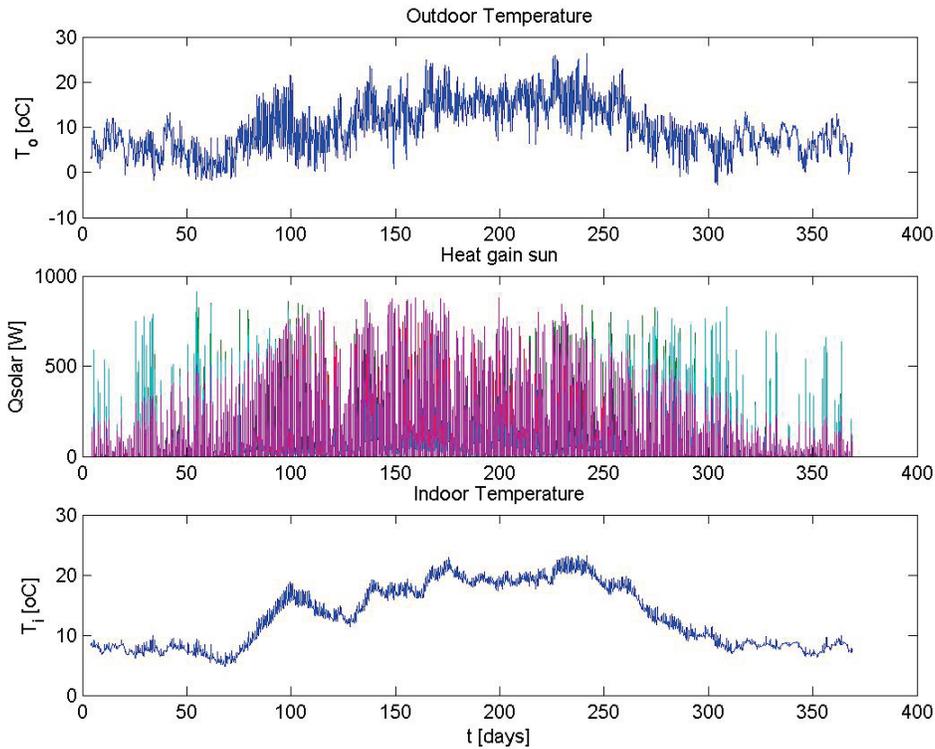

Figure 2: Input variables, outside air temperature and solar gains in one room, and output variable, inside air temperature in one room, derived from a discrete time simulation in HAMBase/Matlab. ($C_{f,T0}$ = 3.09, $C_{f,Qsolar}$ = 5.47, and $C_{f,Ti}$ = 2.61 (Equation (2))).

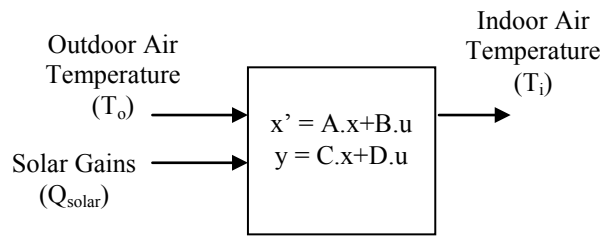

Figure 3: Continuous time state-space model of the building dynamics describing the indoor air temperature in the building. Input variables are the outdoor air temperature and the solar heat gains.

Several system identification models of different mathematical order have been developed. A way of obtaining insight into the quality of a model is to simulate the model with the input from a fresh data set and compare the simulated output with the measured one. This gives a feel for which properties of the system have been picked up by the model, and which have not. Table 2 presents a comparison of the models based on the second half of the original discrete time results from HAMBase/Matlab and a simulation of the SI model (discrete time).

The mean ($\mu_{\varepsilon,I}$) and the standard deviation ($\sigma_{\varepsilon,I}$) of the error have been presented for several models of different mathematical order. The research showed that the mean error ($\mu_{\varepsilon,I}$) is representative for the accuracy of the SI model. Considering the accuracy of the identified models, the standard deviation showed to be of minor importance. Furthermore, Table 3 shows that the accuracy of the model increases with increasing mathematical model order.

Table 3: Comparison of SI models

| Model order | $\mu_{\varepsilon,I}$ [$^o$C] | $\sigma_{\varepsilon,I}$ [$^o$C] |
|---|---|---|
| 1 | 0.9891 | 1.9983 |
| 2 | 0.6215 | 1.1735 |
| 3 | 0.7299 | 0.6956 |
| 4 | 0.2023 | 0.3157 |
| 8 | 0.0488 | 0.2371 |
| 18 | 0.0073 | 0.2356 |
| 40 | 0.0133 | 0.2674 |

(*II, III.*) The research proceeds with a simulation of the reference building in HAMBase/Simulink with free-floating indoor air temperature. A comparison between the results of the SI model simulation (continuous time) and the HAMBase/Simulink simulation with free-floating indoor air temperature (continuous time) is presented in Figure 4. Considering the error between the system identification model and the HAMBase/Simulink model, the figure shows an average error ($\mu_{\varepsilon,II}$) of 0.1868$^o$C. Furthermore, a frequency analysis showed that dynamics with a maximum frequency of approximately $2.10^{-3}$ Hz are present in the SI model. Information regarding the fast dynamic behaviour within one second is not captured by the model. The observation that the frequency of the identified system dynamics is limited to the sample frequency of the signal, used for identification of the model, is confirmed by the Nyquist-Shannon theorem Brigham (2004). The simulation results show that model identification is useful for identification and prediction of free-floating building dynamics for frequencies smaller than the sample frequency.

In conclusion, the research shows that system identification is useful for the identification of free-floating building dynamics and application of the SI model in the same free-floating configuration. The computing time used for the simulation of the SI model in MATLAB is considerably small compared to the computing time needed for the simulation of the HAMBase/Simulink model. The computing time is reduced from several minutes to several seconds. The research proceeds with the identification free-floating building dynamics and the application of the identified model with indoor air temperature control.

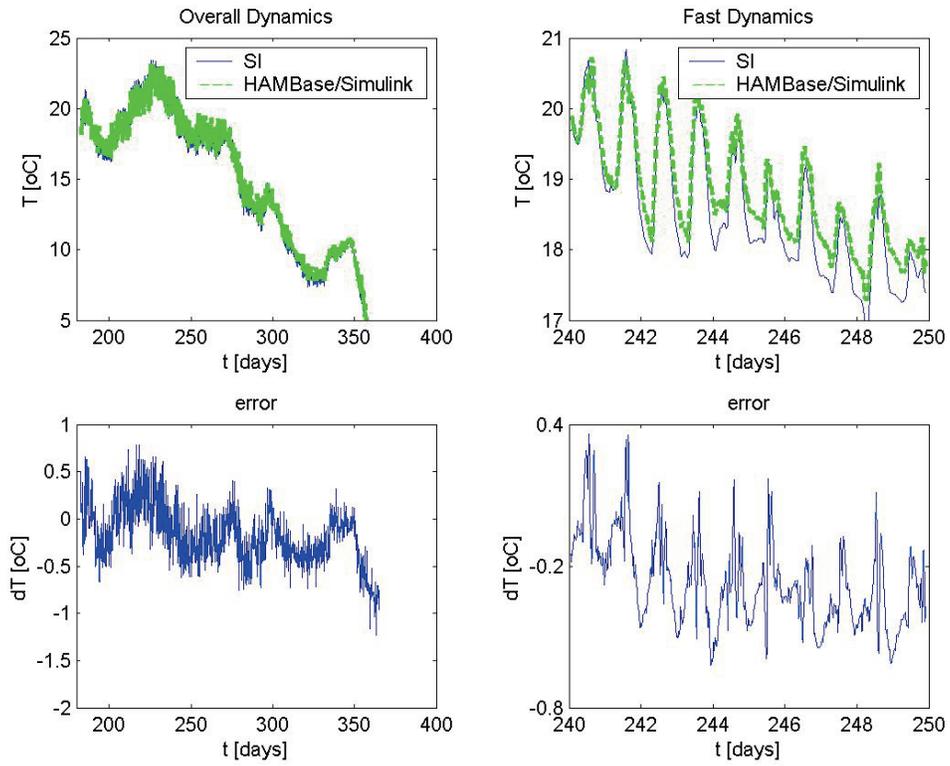

Figure 4: Comparison between the indoor air temperatures predicted by the SI model and the HAMBase/Simulink model.

## 3.2. Free-floating based model identification and climate control application

This section describes the identification of a building model based on discrete time data, containing free-floating indoor air temperature and application in a situation with indoor air temperature control (continuous time). The SI model, which has been identified based on a simulation with free-floating indoor air temperature (Section 3.1), has been simulated with indoor air temperature control in MATLAB/Simulink. (*II.*). The indoor temperature is controlled based on a (continuous) time step, which lies within the time step that has been used for identification of the SI model. Therefore, fast dynamic system behaviour must be present in the SI model to predict the indoor air temperature accurately.

A control strategy based on the SI model combined with on/off-control of the indoor air temperature has been established. Figure 5 illustrates the structure of the building model. Input variables of the SI model are the outside air temperature, solar gains (Figure 3) and the heating and cooling power supplied to the room by the HVAC installation. The heat supplied by the HVAC installation has been added to the solar heat gain and the sum of both is used as a new input for the model. Output variables are the inside air temperature in the building. The HVAC installation is controlled by feedback control of the indoor air temperature.

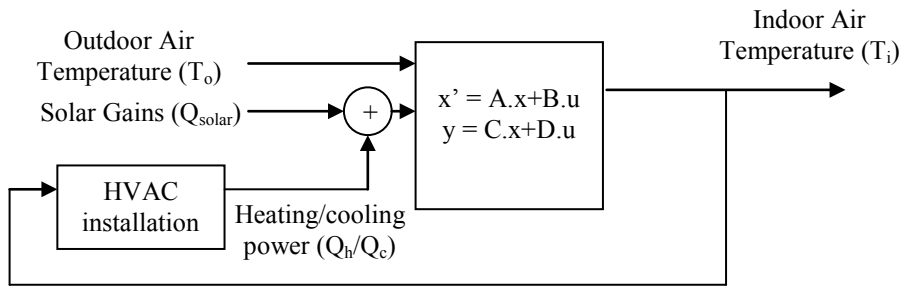

Figure 5: Structure of the building model. Input variables are the outside air temperature, solar gains and the heating and cooling power. The heat supplied by the HVAC installation has been added to the solar heat gain and the sum of both is used as a new input for the model. Output variables are the indoor air temperatures in the building.

(*III.*) Both the SI building model (Figure 5) and the reference HAMBase/Simulink reference model have been simulated with set points for heating and cooling of 18°C and 22°C. The temperature in the building is controlled 24 hours a day. The heating and cooling capacities in the room are respectively 1500 W and 1000 W. The results from both simulations have been compared. Figure 6 shows the indoor air temperature in a room of the building. Considering the error between the results from the SI model simulation and the HAMBase/Simulink results, the figure shows an average error ($\mu_{\varepsilon,II}$) of 0.1 °C and a standard deviation ($\sigma_{\varepsilon,II}$) of 1.76 °C. Moreover, the SI model estimates the yearly energy consumption for heating and cooling for room 1 to be $3.98*10^{10}$ J compared to $3.85*10^{10}$ J predicted by the HAMBase/Simulink model. In conclusion, a relatively large deviation between the building dynamics represented by the SI model and the HAMBase/Simulink building model is present.

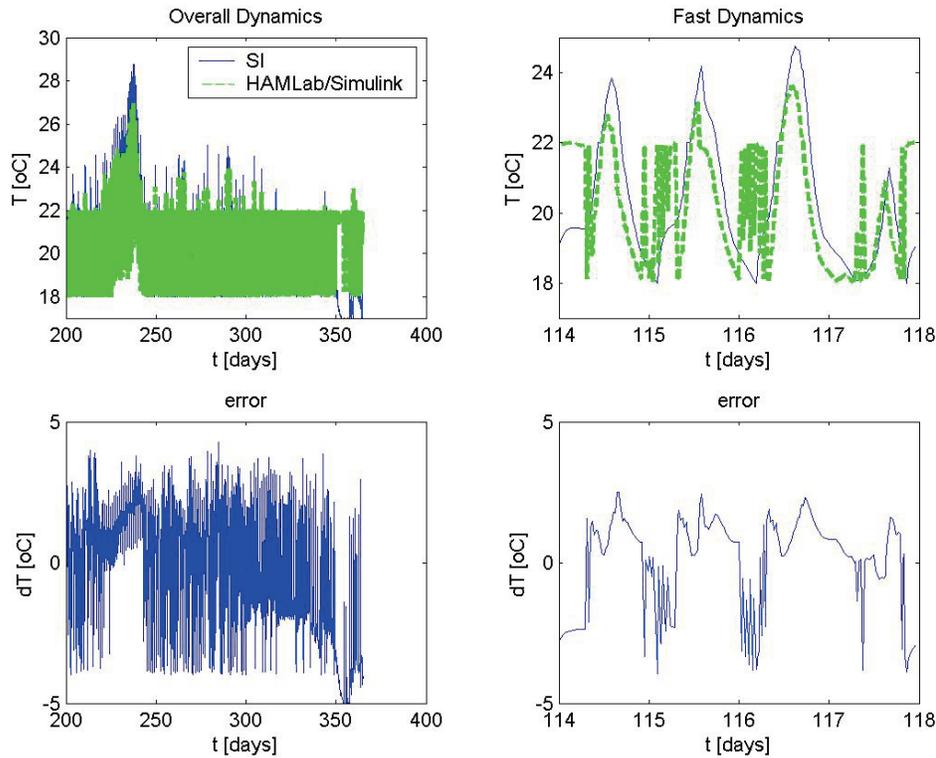

Figure 6: Indoor air temperature in two rooms of the building predicted by the SI model and the HAMBase/Simulink model. Both models have been simulated with set points for heating and cooling of 18°C and 22°C

The research shows that system identification is not applicable for the design of a climate control strategy based on discrete data containing (free-floating) inside air temperature, outside air temperature and solar gains. A frequency analysis shows that system dynamics with a maximum frequency of approximately $2.10^{-3}$ Hz are present in the SI model. Information regarding the fast dynamic behaviour within ten minutes is not captured by the model. The observation that the frequency of the identified system dynamics is limited to the sample frequency of the signal, used for identification of the model, is confirmed by the Nyquist-Shannon theorem (Brigham (2004)). However the average error ($\mu_{\varepsilon,II}$) is acceptable not all properties of the building have been picked up by the SI model, resulting in a relatively large fluctuation of the error. This difference in system behaviour may be caused by the fact that the transfer from heating/cooling power to indoor air temperature is not defined in the model. Moreover, it is assumed that the transfer from heating/cooling power to indoor air temperature is similar to the transfer from the solar gains to the indoor air temperature. After all, this assumption seems invalid.

Often, indoor air temperature in a building as well as outdoor air temperature and solar gains are measured after a building has been built. These data are indirectly used to decide what type of HVAC installation is installed and to develop a proper indoor air temperature control strategy. A system identification model of building dynamics is obtained based on data obtained from measurements in a free-floating situation. However, the reader should notice that, when discrete time data, for example measurements, are used to develop an SI model of building dynamics, this model can only be applied for simulation (1) with a maximum sample frequency as used for identification and (2) with free-floating climate conditions . The model is not suitable for simulation with climate control and control strategy design, since the model lacks (1) valuable information for smaller frequencies than the sample frequency of the input signal used for identification and (2) information regarding the transfer from heating/cooling power to indoor air temperature.

    The research proceeds with the identification of building dynamics based on on/off-control, in order to include the transfer from heating/cooling power to indoor air temperature in the SI model.

## 3.3 On/off-control based model identification (discrete time) and on/off-control based model simulation (discrete time)

This section describes the identification of a building model based on discrete time data, containing on/off-controlled indoor air temperature and application in a situation with a similar type of indoor air temperature control (discrete time). A model based on an on/off-controlled simulation in HAMBase/Matlab is developed. Instead of a free-floating indoor air temperature (Section 3.1), the temperature in the building is controlled by on/off-control of the HVAC installation with set points for heating and cooling of respectively 18$^o$C and 22$^o$C. To these set points, which have been used in the HAMBase/Matlab simulation for identification of the model, is referred to as the *identification set points*. The identification of an SI model based on an on/off-controlled simulation enables the identification of the transfer from heating/cooling power to indoor air temperature. The modelling approach is similar to the approach that is described in Section 3.1.

(*I.*) A simulation time step of one hour is used based on climate data of the years 1974. Input variables of the model are the outside air temperature (Figure 2), the solar gains (Figure 2) and heating/cooling power (Figure 7). The output variables are the inside air temperature in the building. Before identification of building dynamics, the input signal is examined. The sample frequency of the data is $2.8 \cdot 10^{-4}$ Hz. Based on the Nyquist-Shannon theorem, the model is expected to represent building dynamics for frequencies smaller than $2.8 \cdot 10^{-4}$ Hz. The amount of input power into the system is determined by calculation of the Crest factors $C_{f,T0}$, $C_{f,Qsolar}$, $C_{f,Ti}$, and $C_{f,Qh/Qc}$ (Equation (2)). Regarding the outside air temperature, the solar gains and the heating/cooling power the Crest factors are respectively 3.09 ($C_{f,T0}$), 5.47 ($C_{f,Qsolar}$), and 2.23 ($C_{f,Qh/Qc}$). Considering the inside air temperature a Crest factor of 1.76 ($C_{f,Ti}$) and has been calculated. After examination of the input signal, the MATLAB System Identification Toolbox is used to develop a discrete time state-space model describing indoor air temperature. A fourth order state-space model has been selected. Based on the results presented in Table 3, a relatively accurate model is expected with an average error ($\mu_{\varepsilon,I}$) of approximately 0.20.

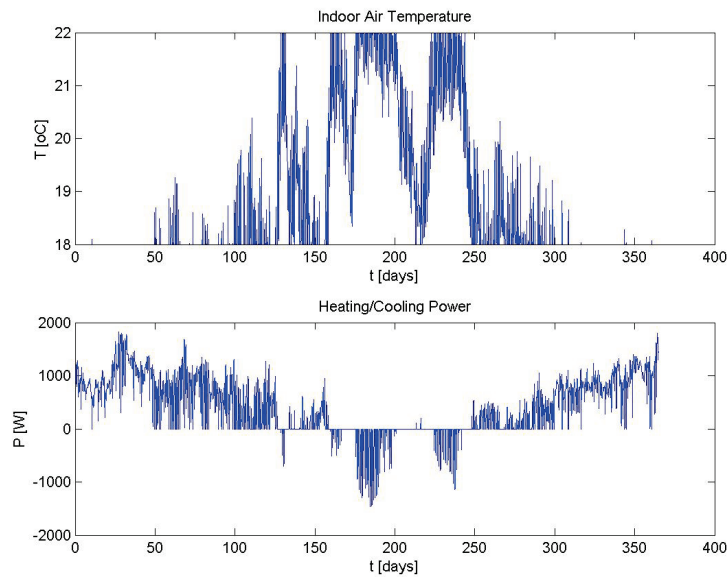

Figure 7: Indoor air temperature and heating/cooling power, derived from a discrete time building simulation in HAMBase/Matlab with indoor air temperature control. For simplicity, heating/cooling power and indoor air temperature of only one room are depicted. ($C_{f,Ti}$ = 1.76 and $C_{f,Qh/Qc}$ = 2.23)

(*II.*) A control strategy based on the fourth order SI model combined with on/off-control of the indoor air temperature has been established. Climate data of the year 2000 has been used for simulation. Input variables of the SI model are the outside air temperature, solar gains and the heating and cooling power supplied to the room by the HVAC installation. Output variables are the indoor air temperatures in the building. The model, which is presented in Figure 8, is implemented in MATLAB and simulated with time steps of one hour.

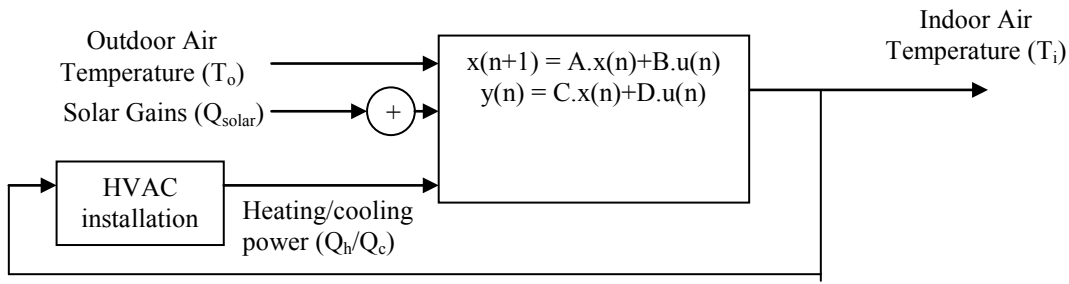

Figure 8: Discrete time SI model combined with on/off-control.

(*III.*) The reference model of the building has been simulated in HAMBase/Matlab. Figure 9 shows that HAMBase/Matlab predicts the indoor air temperature to be between 18°C and 22°C. Comparing the results from the reference model with the results from the SI model simulation, an average error ($\mu_{\varepsilon,II}$) of 0.0533 and a standard deviation ($\sigma_{\varepsilon,II}$) of 0.3971 are observed. Figure 9 shows that the results of both simulations are nearly identical. Comparing the SI model and the reference model, an analysis of dynamic system behaviour (Figure 9) and a frequency analysis show that the dynamic properties of the building within the sample frequency of the input signal have been picked up by the SI model. The SI model contains no information for larger frequencies than the sample frequency.

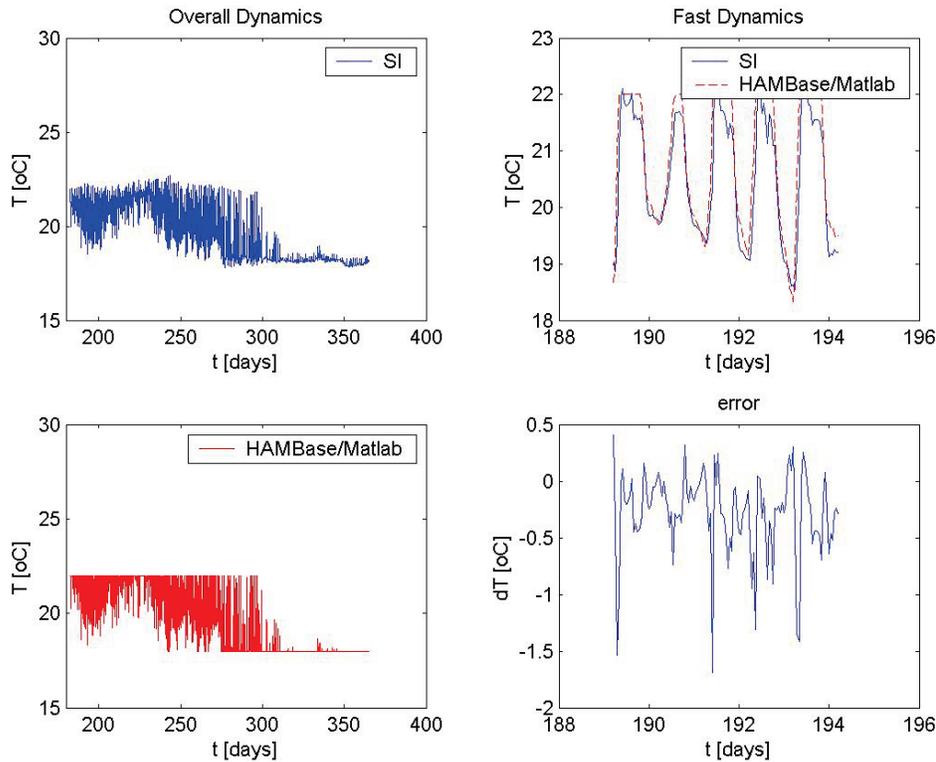

Figure 9: Comparison between indoor air temperatures predicted by the SI model and the HAMBase/Matlab reference model. Both simulations have been performed with set points for heating and cooling of 18°C and 22°C and a time step of one hour.

In conclusion, the research shows that system identification is useful for the identification of building dynamics based on a simulation with on/off-controlled indoor air temperature and application of the SI model in the same on/off-controlled configuration. Moreover, the application of the model is limited to a situation using a similar (discrete) time step, a similar configuration of the building, and similar set points as has been used for model identification. The study proceeds with the research of the applicability of the SI model for continuous time steps as well as for different set points for heating and cooling. The research of the applicability for continuous time steps is reported in Section 3.4. Section 3.5 reports the applicability with respect to different set points for heating and cooling.

## 3.4 On/off-control based model identification (discrete time) and on/off-control based model simulation (continuous time)

This section describes the identification of a building model based on discrete time data, containing on/off-controlled indoor air temperature and application in a situation with a similar type of indoor air temperature control (continuous time). A SI model based on an on/off-controlled simulation in HAMBase/Matlab has been developed. The indoor air temperature in the building is controlled by on/off-control of the HVAC installation with *identification set points* for heating and cooling of respectively 18°C and 22°C. The modelling approach is similar to the approach that is described in Section 3.1.

(*I.*) A simulation time step of one hour is used based on climate data of the years 1974 and 2000. Input variables of the model are the outside air temperature (Figure 2), the solar gains (Figure 2) and heating/cooling power (Figure 7). The output variables are the inside air temperature in the building (Figure 7). Before identification of building dynamics, the input signal is examined. The sample frequency of the data is $2.8 \cdot 10^{-4}$ Hz. Based on the Nyquist-Shannon theorem, the model is expected to represent building dynamics for frequencies smaller than $2.8 \cdot 10^{-4}$ Hz. The amount of input power into the system is determined by calculation of the Crest factors $C_{f,T0}$, $C_{f,Qsolar}$, $C_{f,Ti}$, and $C_{f,Qh/Qc}$ (Equation (2)). Regarding the outside air temperature, the solar gains and the heating/cooling power the Crest factors are respectively 3.09 ($C_{f,T0}$), 5.47 ($C_{f,Qsolar}$), and 2.23 ($C_{f,Qh/Qc}$). Considering the inside air temperature a Crest factor of 1.76 ($C_{f,Ti}$) and has been calculated. After examination of the input signal, the MATLAB System Identification Toolbox is used to develop a continuous time state-space model describing indoor air temperature. A fourth order state-space model has been selected. Based on the results presented in Table 2, a relatively accurate model is expected with an average error ($\mu_{\varepsilon,I}$) of approximately 0.20.

(*II.*) A control strategy based on the fourth order SI model combined with on/off-control of the indoor air temperature has been established. Input variables of the SI model are the outside air temperature, solar gains and the heating and cooling power supplied to the room by the HVAC installation. Output variables are the indoor air temperatures in the building. The model, which is presented in Figure 10, is implemented in MATLAB/Simulink.

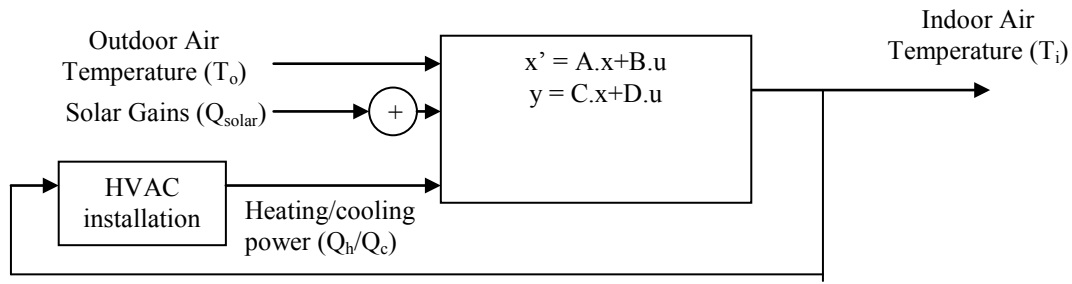

Figure 10: Continuous time SI model combined with on/off-control.

(*III.*) The reference model of the building has been simulated in HAMBase/Simulink. Figure 11 shows that HAMBase/Simulink predicts the indoor air temperature to be between 18°C and 22°C. Comparing the results from HAMBase/Simulink with the results from the SI model simulation, an average error ($\mu_{\varepsilon,II}$) of 0.25 and a standard deviation ($\sigma_{\varepsilon,II}$) of 1.99 have been observed. Focusing on the cooling set point, the SI model estimates the indoor air temperature to be considerably higher compared to the temperature predicted by the HAMBase/Simulink model, although both models have been simulated with the same capacities of heating and cooling power. Fast dynamic properties have not been picked up by the SI model.

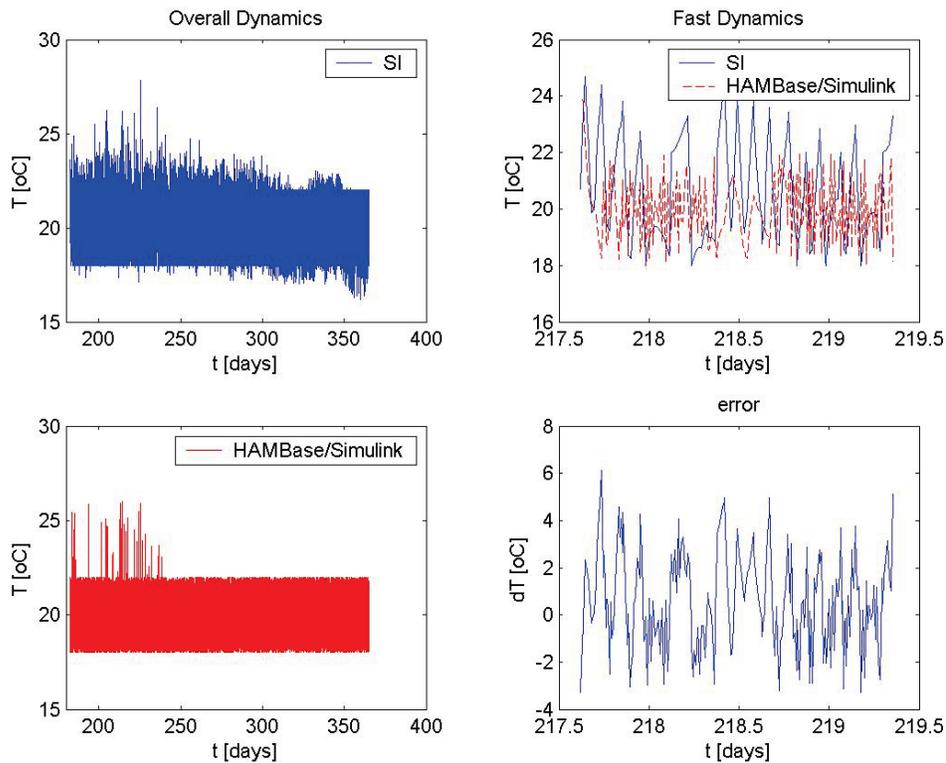

Figure 11: Comparison between indoor air temperatures predicted by the SI model and the HAMBase/Simulink model. Both simulations have been performed with set points for heating and cooling of 18°C and 22°C.

Considering overall system dynamics both the results of the SI model and the HAMBase/Simulink reference model agree well with each other. The mean error of the results ($\mu_{\varepsilon,I}$ and ($\mu_{\varepsilon,II}$) is relatively small. Regarding fast building dynamics within the identification time step of 1 hour, system dynamics has not been captured by the SI model. Figure 11 shows that fast dynamic behaviour of the building within one hour (sample time) has not been picked up by the SI model. Information regarding system dynamics within the identification time step of one hour is not present in the SI model.

The research shows that an SI model that has been identified based on a discrete time step of one hour is not applicable for simulation in continuous time. Moreover, it is concluded that system identification is not suitable for the identification of fast building dynamics within the time step that has been used for identification of the model. This observation is confirmed theoretically by the Nyquist-Shannon theorem.

The research proceeds with the study of system identification with respect to the simulation of an SI model with different set points for heating and cooling than have been used for identification (*identification set points*).

## 3.5 On/off-control based model identification (on/off-control A) and on/off-control based model simulation (on/off-control B)

This section describes the identification of a building model based on discrete time data, containing on/off-controlled indoor air temperature and application in a situation with a different type of indoor air temperature control (discrete time). The sensitivity of the *identification set points* with respect to the transfer information between SI model output variable (indoor air temperature) and controlled input variable (heating/cooling power) has been researched. First, building energy simulations in HAMBase/Matlab have been performed using different *identification set points* for heating and cooling, resulting in several discrete time data sets (*I.*). Second, for every data set, the input signal $T_i$ has been examined focusing on the amount of input power present in the input signal ($T_i$). Therefore, the corresponding Crest factor $C_{f,Ti}$ (Equation (2)) has been calculated. Third, system identification models have been identified based on these data sets. Lastly, the SI models have been simulated in MATLAB (*II.*) (discrete time) and the results have been compared (*III.*) with the simulation results of the HAMBase/Matlab reference model. Both simulations have been performed with on/off-control between 18°C and 22°C.

A comparison of the results of the SI models with both the results of a reference simulation, using the original data set and specific *identification set points* ($\mu_{\varepsilon,I}$), and the results obtained from a reference simulation, using a new data set and set points for heating and cooling of 18°C and 22°C ($\mu_{\varepsilon,II}$) have been presented in Table 4 For every SI model, Table 4 resents the *identification set points* that have been used for identification, the Crest factor with respect to the indoor air temperature $T_i$, and the mean error of the results ($\mu_{\varepsilon,I}$) and ($\mu_{\varepsilon,II}$).

Table 4: Comparison between on/off-controlled simulations

| Model | Order | Identification set points | | Crest Factor $C_{f,Ti}$ | Mean error I ($\mu_{\varepsilon,I}$) [$^{o}$C] | Mean error II ($\mu_{\varepsilon,II}$) [$^{o}$C] | Std error II |
|---|---|---|---|---|---|---|---|
| | | $T_h$ | $T_c$ | | | | |
| SI | 4 | 10 | 24 | 1.9152 | 0.0063 | 0.1978 | 0.6231 |
| SI | 8 | 10 | 24 | 1.9152 | 0.0067 | 0.2992 | 0.2986 |
| SI | 24 | 10 | 24 | 1.9152 | 0.0385 | 0.3198 | 0.3168 |
| SI | 4 | 12 | 24 | 2.0702 | 0.3430 | 0.269 | 0.3362 |
| SI | 8 | 12 | 24 | 2.0702 | 0.0138 | 0.2266 | 0.2658 |
| **SI** | **4** | **16** | **24** | **2.4962** | **0.0334** | **0.09058** | **0.3061** |
| **SI** | **8** | **16** | **24** | **2.4962** | **0.0089** | **0.1073** | **0.1455** |
| SI | 4 | 16 | 22 | 2.1682 | 0.0591 | 0.1341 | 0.3858 |
| **SI** | **8** | **16** | **22** | **2.1682** | **0.0144** | **0.03871** | **0.1535** |
| SI | 4 | 18 | 22 | 2.3863 | 0.0656 | 0.1429 | 0.506 |
| **SI** | **8** | **18** | **22** | **2.3863** | **0.0371** | **0.08719** | **0.1641** |
| SI | 4 | 20 | 22 | 2.5912 | 0.0432 | 0.2115 | 0.573 |
| **SI** | **8** | **20** | **22** | **2.5912** | **0.0102** | **0.08155** | **0.1529** |
| SI | 4 | 21 | 22 | 6.9098 | 0.0053 | 1.693 | 2.901 |
| SI | 8 | 21 | 22 | 6.9098 | 0.0125 | 6.048 | 3.775 |
| SI | 4 | 20 | 26 | 3.6785 | 0.0430 | 0.07722 | 0.4262 |
| **SI** | **8** | **20** | **26** | **3.6785** | **6.1770e-4** | **0.118** | **0.1209** |
| SI | 4 | 16 | 26 | 2.8896 | 0.0218 | 0.2362 | 0.3097 |
| **SI** | **8** | **16** | **26** | **2.8896** | **0.0340** | **0.0838** | **0.165** |
| SI | 4 | 5 | 26 | 1.9365 | 0.2367 | 1.305 | 2.072 |
| SI | 8 | 5 | 26 | 1.9365 | 0.0085 | 0.7084 | 0.9323 |

The identified models have been analyzed focussing on overall system dynamics and on fast system dynamics. Considering overall system dynamics, models have been selected based on the criterion that provided the mean error ($\mu_{\varepsilon,II}$) between the results of the SI model simulation and the HAMBase/Matlab reference model is smaller than 0.12, relatively accurate results have been obtained by the SI model simulations. The models that fulfil the criterion have been bolded in Table 4. Moreover, Table 4 shows that the accuracy of a model is both dependent of the mathematical order and the *identification set points* that have been used for simulation. First of all, the influence of the mathematical order on model accuracy is considered. Considering the mean error ($\mu_{\varepsilon,I}$) between the SI model simulation results and the validation data, the accuracy of the model increases with increasing mathematical order. The research showed that the accuracy of a model can be determined by examining mean error I ($\mu_{\varepsilon,I}$).

Second, the influence of the *identification set points* on model accuracy is considered. Table 4 shows that when the input signal contains insufficient input power to excite the system, the SI model contains insufficient transfer information regarding input-output behaviour. This lacking information results in an inaccurate prediction of the indoor air temperature ($\mu_{\varepsilon,II}$). For example, the results of an SI model, which has been identified based on *identification set points* of 21°C for heating and 22°C for cooling, are relatively inaccurate ($\mu_{\varepsilon,II}$ = 6.948). The relatively high Crest factors $C_{f,Ti}$ show that the input signal contains little input power.

In Figure 12, the indoor air temperatures predicted by the SI model (based on *identification set points* of 21°C and 22°C) and the HAMBase/Matlab reference model are presented. Both models have been simulated with set points for heating and cooling of 18°C and 22°C. The SI model contains insufficient transfer information regarding input-output behaviour, resulting in an inaccurate prediction of the indoor air temperature. Moreover, a considerable deviation between the dynamics represented by the SI model and the reference model is present. The example shows that however an accurate model is expected (mean error I ($\mu_{\varepsilon,I}$) is relatively small), the model is inaccurate because the input power into the system is too low.

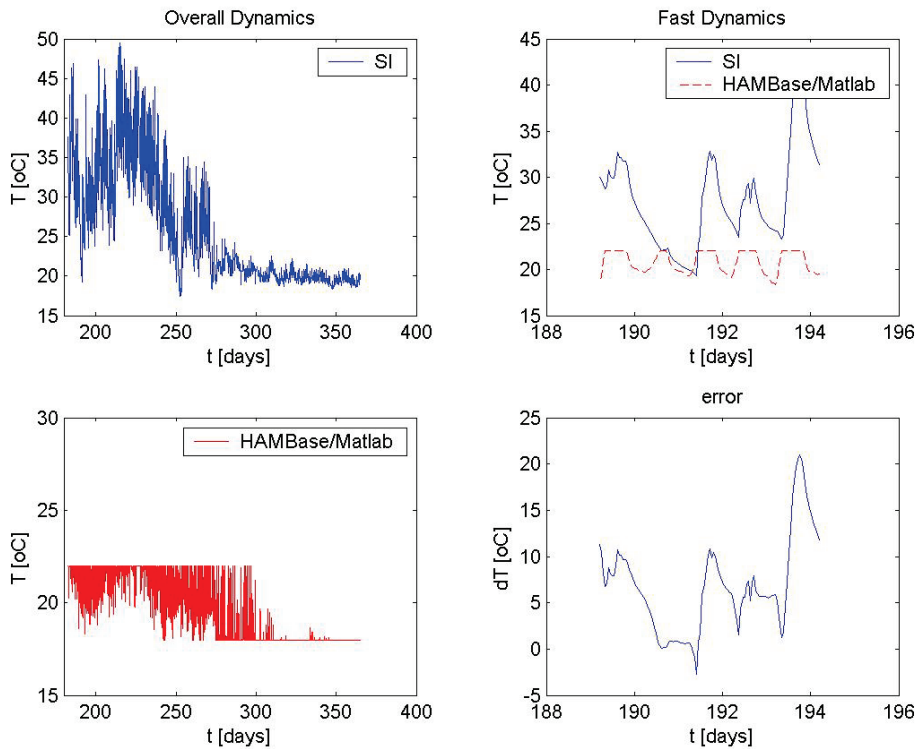

Figure 12: Comparison between indoor air temperatures predicted by the SI model and the HAMBase/Matlab reference model. The SI model has been identified with *identification set points* of 21°C and 22°C and simulated with set points of 18 °C and 22 °C.

The research shows that an SI model that has been identified based on *identification set points* $T_{h,ID}$ and $T_{c,ID}$ is not generally applicable for simulation with different set points for heating and cooling. Table 4 illustrates that for this specific type of building the *identification set point* for heating is limited to a range between 16 °C and 20 °C as well as the *identification set point* for cooling is limited to a range between 22 °C and 26 °C.

In conclusion, system identification is useful for the prediction of overall building dynamics (dynamic behaviour of the building within the identification time step) and control of overall building dynamics. The accuracy of the derived SI model is dependent of both the choice of *identification set points* and the mathematical order of the derived model. In addition, the main advantage is a reduction in computing time and the possibility of performing a simulation in a MATLAB environment, which makes model coupling to other building models and tools possible. However, system identification is not applicable for the identifying of fast building dynamics, which lie within the identification time step and can not be used for indoor air temperature control within this time step.

These observations result in the following guidelines for the successful application of system identification models for the prediction of overall dynamics. First of all, before identification of the model the input signal should be examined, focusing on the sample frequency as well as the input power present in the input signals. The calculation of the Crest factor (Equation (2)) is a good way of giving insight into the input power. Second, the SI model is identified and simulated. Third, the simulation results are compared with the original data set, based on mean error I ($\mu_{\varepsilon,I}$). Provided the model contains sufficient transfer information regarding input-output behaviour ($C_f < 4$) and the model is relatively accurate ($\mu_{\varepsilon,I} < 0.05\ ^oC$), the model is suitable for the prediction of the indoor air temperature in a building.

The research shows that system identification offers good perspectives for the prediction of the overall dynamical behaviour regarding the indoor air temperature in a building. The research proceeds with the analysis of HAM processes. The inclusion of the dynamical behaviour regarding the moisture contents in the building is reported in Section 4.

## 4. HAM MODEL

Section 3 describes the identification of building dynamics focusing on indoor air temperature. This section focuses on the application of system identification regarding Heat, Air and Moisture (HAM) transport. A system identification model is used to predict indoor air temperature as well as indoor air humidity. The application for temperature and humidity control has been researched.

A similar modelling methodology as depicted in Figure 1 is applied. First, (*I.*) HAMBase/Matlab is used to perform a HAM simulation with time steps of one hour, based on climate data of the year 1976. *Identification set points* for heating and cooling of 18$^o$C ($T_{h,ID}$) and 22$^o$C ($T_{c,ID}$) and *identification set points* for humidification and dehumidification of 40% ($RH_{hum,ID}$) and 70% ($RH_{dehum,ID}$) relative humidity. Models' input and output variables are the indoor and outdoor air temperature, the indoor and outdoor air relative humidity and the power supplied to the HVAC installation. The input and output variables are presented in Figure 13. Before identification of building dynamics, the input signal is examined. The sample frequency of the data is $2.8 \cdot 10^{-4}$ Hz. Based on the Nyquist-Shannon theorem, the model is expected to represent building dynamics for frequencies smaller than $2.8 \cdot 10^{-4}$ Hz. The amount of input power into the system is determined by calculation of the Crest factors $C_{f,To}$, $C_{f,RHo}$, $C_{f,Qsolar}$, $C_{f,Ti}$, $C_{f,RHi}$, $C_{f,Qh/Qc}$, and $C_{f,Q(de)hum}$ (Equation (2)). The Crest factors are respectively 3.09 ($C_{f,T0}$), 5.47 ($C_{f,Qsolar}$), 5.76 ($C_{f,Rho}$), 1.71 ($C_{f,Ti}$), 2.39 ($C_{f,RHi}$), 2.16 ($C_{f,Qh/Qc}$), and 5.28 ($C_{f,Q(de)hum}$). Second, an SI model has been identified based on the simulation results obtained from HAMBase/Matlab. Third, the SI model has been simulated in MATLAB using climate data of the year 2000 and with set points of 18$^o$C ($T_h$) for heating, 22$^o$C ($T_c$) for cooling, 40% ($RH_{hum}$) for humidification and 70% ($RH_{dehum}$) for dehumidification.

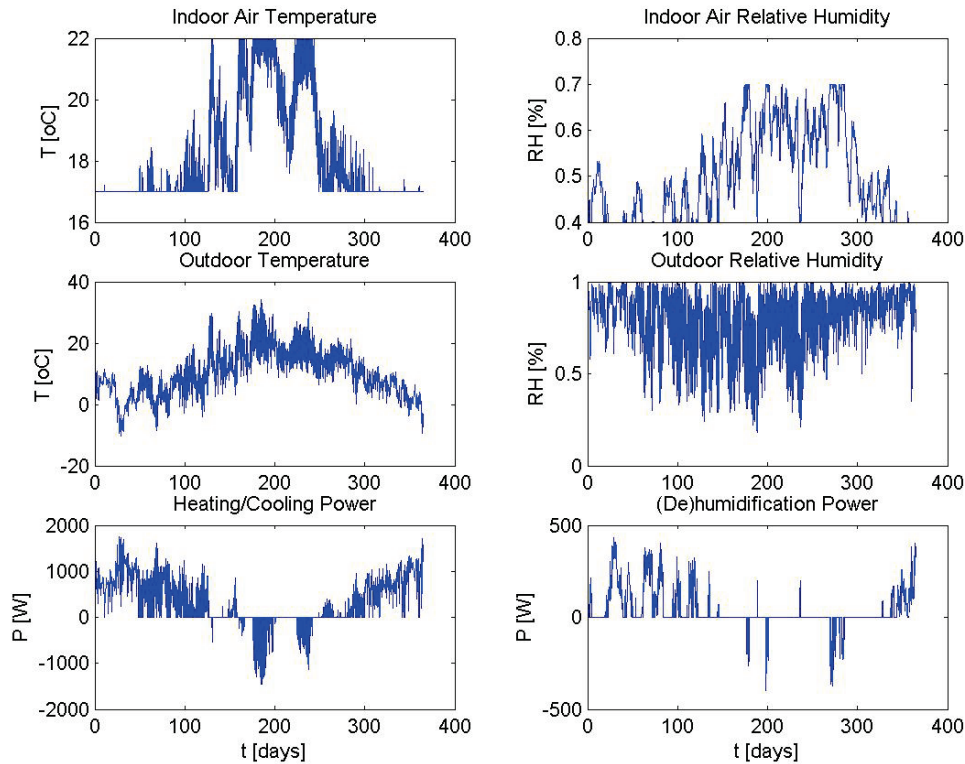

Figure 13: Input and output variables of the system identification model: the air temperature, the air relative humidity and the power supplied to the HVAC installation.

The HAM model of the building with feedback control is shown in Figure 14. Both indoor air temperature and indoor air humidity are controlled by the HVAC installation. The model has been simulated in MATLAB with time steps of one hour (II.) and compared with the simulation results of the HAMBase/Matlab reference model.

The simulation results of both the SI model and the reference model have been presented in Figures 15 to 17. Figure 15 shows the indoor air temperature in the building predicted by the SI model and the reference model (HAMBase/Matlab). Figure 16 shows the indoor air absolute humidity. The relative humidity of the air in the building is presented in Figure 17. Comparing overall dynamics, an average error of respectively $0.24^{\circ}C$ ($\mu_{II,T}$), $0.21.10^{-5}$ kg/kg ($\mu_{II,X}$) and 4.2% RH ($\mu_{II,RH}$) is observed. A spectral analysis shows that both the SI model results and the results obtained from HAMBase/Matlab agree well with each other. However, the average error of the predicted indoor air relative humidity ($\mu_{II,RH}$) is relatively high. The error is caused by the numerical calculation of the relative indoor air humidity, which is based on the predicted indoor air temperature and absolute humidity. The error in of the indoor air relative humidity is a summation of the error of the indoor air temperature and the absolute humidity. Therefore, the error of the predicted indoor air relative humidity gives no insight in the quality and accuracy of the developed SI model.

In conclusion, the research shows that system identification offers good perspectives for the modelling of the HAM processes in a building. The simulation of an SI model in MATLAB enables the designer of building and HVAC installation to predict the indoor air temperature as well as indoor air humidity. In addition, main advantages are that the building model runs in a MATLAB environment and computing time is reduced. First of all, the implementation of the building model in an MATLAB environment offers the user the possibility of coupling the building model to other components and simulation modules. Second, the simulation of the SI model of a building in MATLAB is less time consuming compared to the simulation of the building model in standard simulation software (for example ESP-r). Focusing on the advantages of the MATLAB environment, the next section describes the identification of a building model based on discrete time data obtained from an ESP-r building simulation.

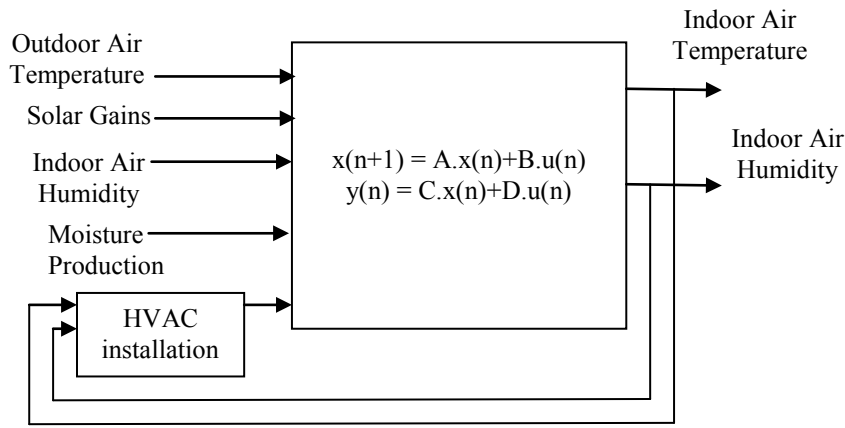

Figure 14: HAM model of the building with feedback control. Indoor air temperature and indoor air humidity are controlled by the HVAC installation.

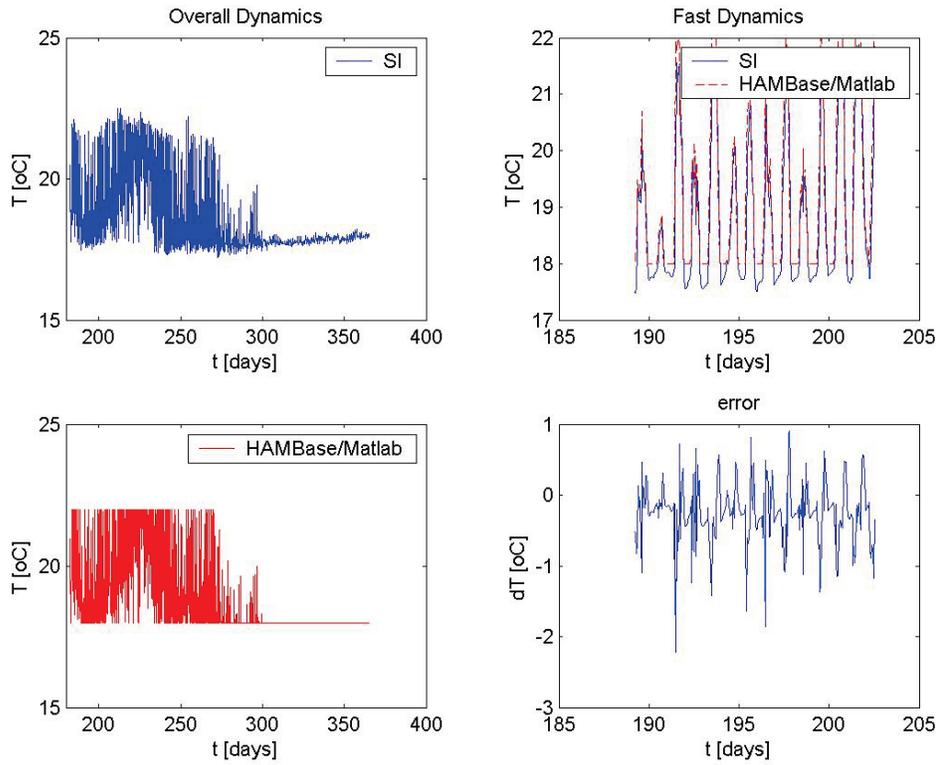

Figure 15: Comparison between indoor air temperature predicted by the HAMBase/Matlab reference model simulation and the results from the SI model simulation ($\mu_{II,T}$ = 0.24°C; $\sigma_{II,T}$ = 0.33°C).

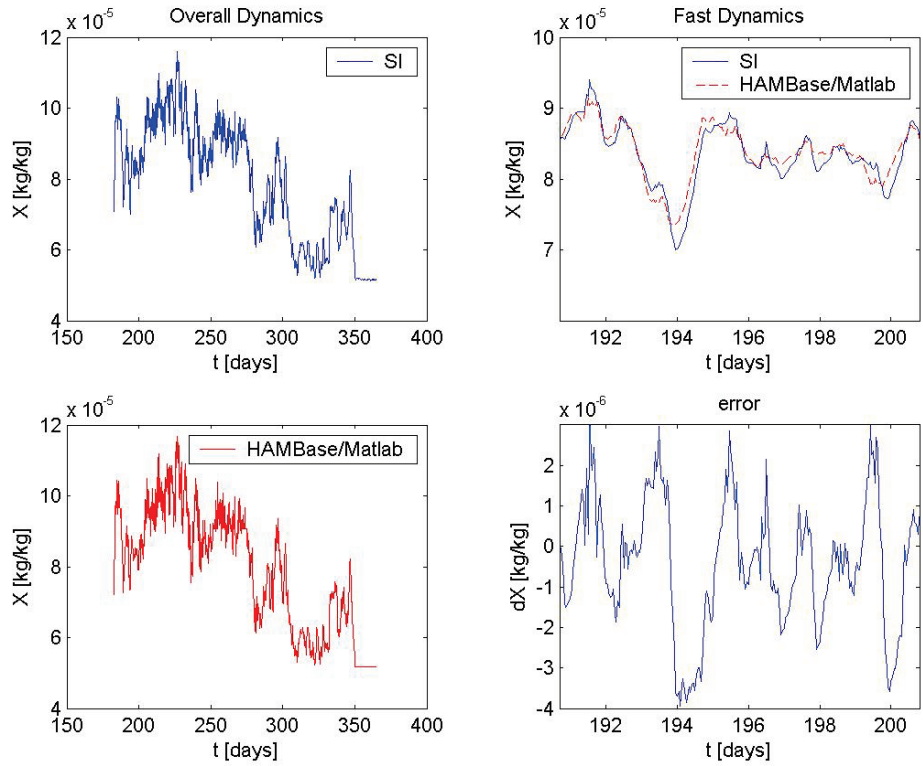

Figure 16: Comparison between indoor air absolute humidity predicted by the HAMBase/Matlab reference model simulation and the results from the SI model simulation ($\mu_{II,X} = 0.21 \cdot 10^{-5}$ kg/kg; $\sigma_{II,X} = 0.28 \cdot 10^{-5}$ kg/kg).

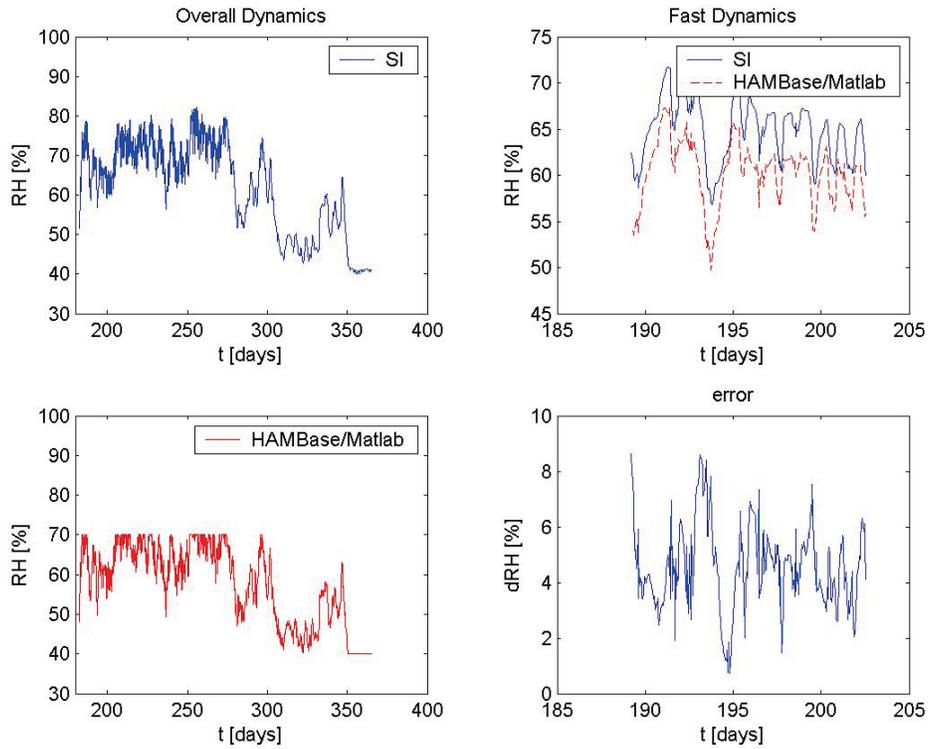

Figure 17: Comparison between indoor air relative humidity predicted by the HAMBase/Matlab reference model simulation and the results from the SI model simulation ($\mu_{II,RH}$ = 4.2% RH; $\sigma_{II,RH}$ = 2.7% RH).

.

## 5. MODEL IDENTIFICATION BASED ON IMPLEMENTED EXTERNAL DATA

This section describes the modeling of building dynamics based on a building performance simulation in ESP-r. First, the indoor air temperature in an office building, consisting of an office, a reception, and a roof space, is simulated in ESP-r. The office building is standard available in the ESP-r library and the geometry of the building is presented in Figure 18. Building dynamics have been simulated for one year with a discrete time step of one hour and set points for heating and cooling of 12$^o$C and 20$^o$C. Second, the results have been exported and input and output variables have been analyzed, focusing on the input power of the system. Input and output variables of the SI model are depicted in Figure 19. The Crest factors ($C_f$) with respect to the outdoor air temperature, the solar gains, the internal heat gains, the heating/cooling power, and the indoor air temperature are respectively 3.21($C_{f,To}$), 5.91 ($C_{f,Qsolar}$), 1.90 ($C_{f,Qint}$), 4.54 ($C_{f,Qh/Qc}$), and 2.82 ($C_{f,Ti}$). The MATLAB System Identification Toolbox has been used for model identification. Since all Crest factors are relatively small, it is expected that sufficient input power is delivered into the system for identification of an accurate SI model. Third, the obtained model SI is used for the design of climate control and a similar control strategy as depicted in Figure 8 has been developed. Lastly, the building model is implemented and simulated in MATLAB.

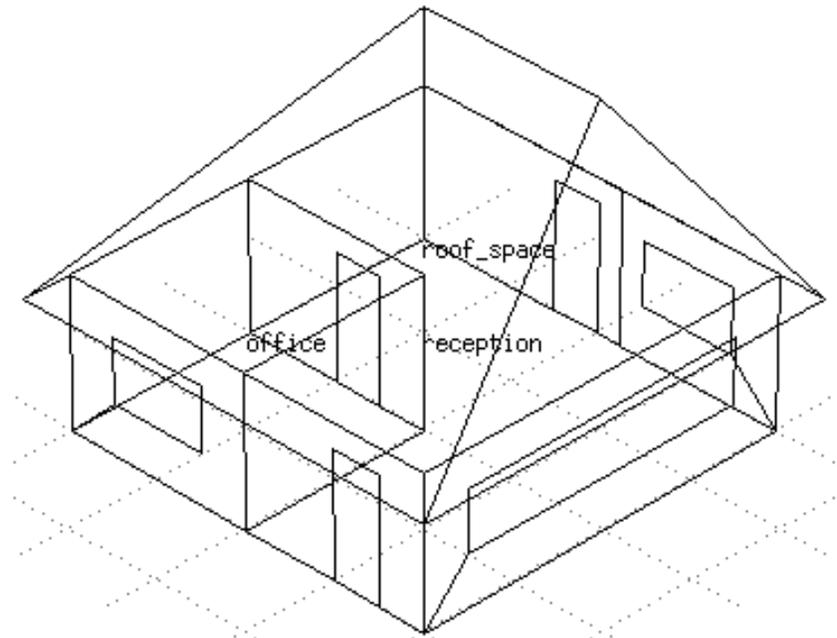

Figure 18: ESP-r office building consisting of an office, a reception and a roof space.

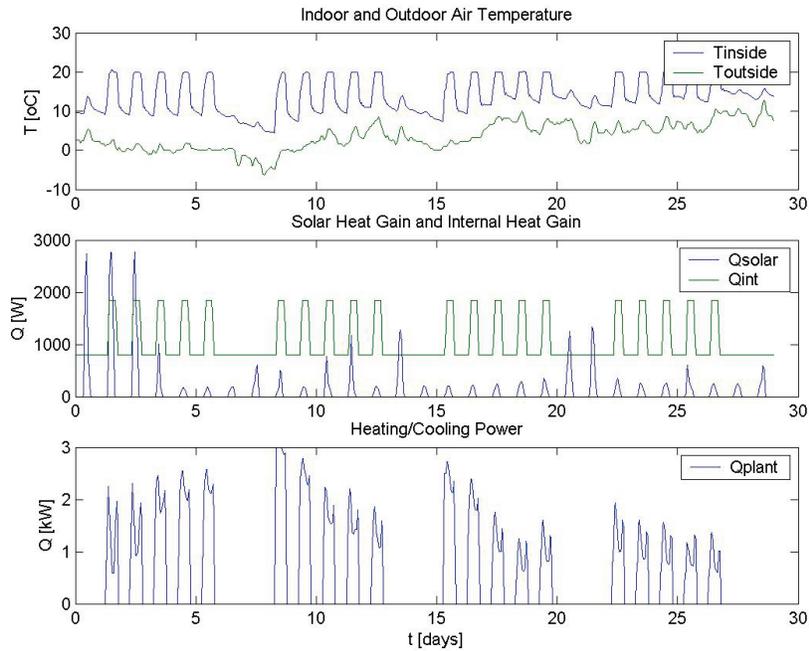

Figure 19: Input and output variables of the SI model.

The results of the simulation of the SI building model in MATLAB are shown in Figure 20. The building model has been simulated with set points for heating and cooling of 12°C and 20°C. Considering overall dynamics, the SI model predicts the indoor air temperature to be for approximately 96% of the time within the set point range. Comparing both models of building dynamics, the average error of the predicted indoor air temperature is 0.13°C ($\mu_{II,T}$). The research shows that the SI model is applicable for the obtaining of information regarding building dynamics. The computing time for the simulation of the SI model of the building is less than a minute. This is considerably fast compared to the computing time of approximately 15 minutes, needed for the ESP-r building simulation. The reduction of computing time offers perspectives for the simulation of large buildings using an SI model.

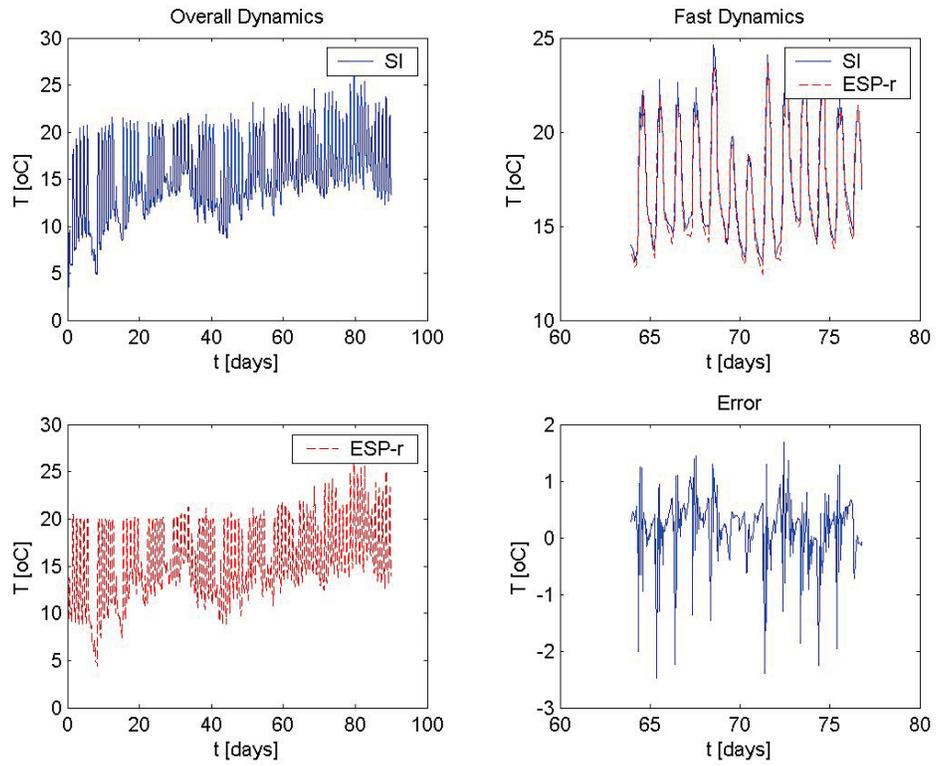

Figure 20: Comparison between the results obtained by ESP-r simulation and the results from the SI model simulation. ($\mu_{II,T}$ = 0.13$^{o}$C; $\sigma_{II,T}$ = 0.98$^{o}$C)

# 6. CONCLUSIONS AND DISCUSSION

This study reports the research of the applicability of system identification for identification of building dynamics and climate control. Considering building simulation, the main problems are that a simulation of a large complex building using standard building simulation software is time consuming and simulation results often lack information regarding fast dynamic behaviour (in the order of seconds) of the building, since most software uses a discrete time step. The applicability of system identification to reduce computing time for the simulation of large complex buildings as well as the applicability of system identification to identify building dynamics based on discrete time data (one hour) for climate control design has been researched.

It is concluded that system identification is applicable for the identification of building dynamics with a frequency that is smaller as the maximum sample frequency as used for identification. System identification offers good perspectives for climate control design, but the application is limited. A summary of the observed limitations of the applicability of system identification with respect to the researched case studies is presented in Table 5.

Table 5: Case studies and observed limitations

| Case Study | Input (Identification) | Output (Application) | Possible | Limitations |
|---|---|---|---|---|
| I | Discrete free-floating indoor air temperature | Continuous free-floating indoor air temperature | Yes | |
| II | Discrete free-floating indoor air temperature | Continuous and discrete controlled indoor air temperature | No | Lacking input-output transfer information |
| III | On/off controlled discrete indoor air temperature | On/off controlled discrete indoor air temperature | Yes | Crest factor |
| IV | On/off controlled discrete indoor air temperature | On/off controlled continuous indoor air temperature | No | Time step, lacking fast dynamics. |
| V | On/off controlled discrete indoor air temperature | On/off controlled discrete indoor air temperature | Yes | Crest factor |

The study showed that:
- o System identification is useful for the identification of free-floating building dynamics and application of the SI model in the same free-floating configuration.
- o System identification is not applicable for the design of a climate control strategy based on discrete data containing (free-floating) inside air temperature, outside air temperature and solar gains. The research showed that the SI model lacks (1) valuable information for smaller frequencies than the sample frequency of the input signal used for identification and (2) information regarding the transfer from heating/cooling power to indoor air temperature.

- System identification is useful for the identification of building dynamics based on a simulation with on/off-controlled indoor air temperature and application of the SI model in the same on/off-controlled configuration. Moreover, the application of the model is limited to a situation using a similar (discrete) time step, a similar configuration of the building, and similar set points as has been used for model identification. With respect to the *identification set points*, a value of the Crest factor between 1 and 4 ($C_{f,Ti} < 4$) is required to enable the identification of an accurate SI model.
- An SI model that has been identified based on a discrete time step of one hour is not applicable for simulation in continuous time. The research showed that it is not possible to capture fast dynamics within the time step that is used for identification of the SI model.
- An SI model that has been identified based on *identification set points* $T_{h,ID}$ and $T_{c,ID}$ is not generally applicable for simulation with different set points for heating and cooling. Therefore, examination of the input signal is required, focusing on the input power present in the input signals. The calculation of the Crest factor (Equation (2)) is a good way of giving insight into this waveform property. Provided the model contains sufficient transfer information regarding input-output behaviour ($C_{f,Ti} < 4$) and the model is relatively accurate ($\mu_{\varepsilon,I} < 0.05\ ^{o}C$), the model is suitable for the prediction of overall building dynamics.

Furthermore, the research shows that system identification offers good perspectives for the modelling of HAM processes in a building. The simulation of an SI model in MATLAB enables the designer of a building and HVAC installation to predict the indoor air temperature as well as the indoor air humidity.

The research shows that the main advantages of system identification models compared to the modelling of building dynamics and the design of climate control using standard building simulation software are, that:

- SI models run in a MATLAB environment, in which many building simulation tools have been developed. This offers good perspectives regarding the coupling of the SI model to other building models, installation components and simulation modules (SIMBAD, COMSOL Multiphysics).
- The computing time regarding the simulation of an SI model is reduced significantly compared to the computing time needed for the simulation of a similar building model using standard building simulation software.

# NOMENCLATURE

| a, b, c, d, e, f | coefficients of the polynomial functions $z^{-1}$ |
|---|---|
| $z^{-1}$ | backward shift operator |
| A, B, C, D | matrixes of the state-space system |
| t | time [s] |
| y | output variable of the state-space system |
| u | input variable of the state-space system |
| x | state variable of the state-space system |
| x' | time derivative of state variable x |
| n | time step |
| T | temperature [$^o$C] |
| $T_i$ | indoor air temperature [$^o$C] |
| $T_o$ | outside air temperature [$^o$C] |
| $T_h$ | set point for heating [$^o$C] |
| $T_c$ | set point for cooling [$^o$C] |
| $T_{h,ID}$ | identification set point for heating [$^o$C] |
| $T_{c,ID}$ | identification set point for cooling [$^o$C] |
| X | air humidity [kg/kg] |
| $X_i$ | indoor air humidity [kg/kg] |
| $X_o$ | outside air humidity [kg/kg] |
| RH | air relative humidity [%] |
| $RH_i$ | indoor air relative humidity [%] |
| $RH_o$ | outside air relative humidity [%] |
| p | electricity consumption [kW] |
| $Q_{solar}$ | solar gains, total heat gain of the solar radiance supplied to the building [W] |
| $Q_h$ | heating power [W] |
| $Q_c$ | cooling power [W] |
| $Q_{(de)hum}$ | (De)humidification power [W] |
| ε | error [$^o$C] |
| P(T) | distribution P of the error T |
| P(ε) | distribution P of the error ε |
| μ | mean |
| σ | standard deviation |
| $C_f$ | Crest factor |
| $C_{f,i}$ | Crest factor with respect to variable i. |
| *Subscripts* | |
| SI | predicted by SI model |

# REFERENCES

Cunningham, M.J., 2001, Inferring ventilation and moisture release rates from field psychometric data only using system identification techniques, *Building and Environment*, vol. 36, no. 1, p. 129-138.

Lowry, G., Lee, M.-W., 2004, Modelling the passive thermal response of a building using sparse BMS data, *Applied Energy*, vol. 78, no. 1, p. 53-62.

Madsen, H., Holst, J., 1995, Estimation of continuous-time models for the heat dynamics of a building, *Energy and Buildings*, vol. 22, no. 1, p. 67-79.

Mechaqrane, A., Zouak M., 2004, A comparison of linear and neural network ARX models applied to a prediction of the indoor temperature of a building, *Neural Computing & Applications*, vol. 13, no. 1, p. 32–37.

Pape, F.L.F., Mitchell, J.W., Beckman W.A., 1991, Optimal control and fault detection in heating, ventilating and air-conditioning systems. *ASHRAE Transactions*, vol. 97, no. 1, p. 729-736.

Crawley, D.B., Hand, J.W., Kummert, M., Griffith, B.T., 2005, Contrasting the capabilities of building energy performance simulation programs, U.S. Department of Energy, Washington, D.C., U.S.A.

ASHRAE Handbook - Fundamentals, 2005, American Society of Heating, Refrigerating and Air-Conditioning Engineers, Atlanta.

Brigham, E.O., 2004, *The Fast Fourier Transform and its applications*, Prentice Hall.

Girod, B., Rabenstein, R., Stenger, A., 2001, *Signals and Systems*, Wiley.

Ljung, L., 1997, *MATLAB System identification toolbox user's guide*, Natick (MA), MathWorks.

De Wit, M.H., 2004, *HAMBase A model for the simulation of the thermal and hygric performance of building and systems*, Technische Universiteit Eindhoven. http://sts.bwk.tue.nl/hambase/.

*Data Model Summary ESP-r Version 9 Series*, 2001, Energy Systems Research Unit, University of Strathclyde. http://www.esru.strath.ac.uk/